\documentclass[twocolumn,showpacs,preprintnumbers,amsmath,amssymb,superscriptaddress]{revtex4}

\usepackage{graphicx}
\usepackage{epsfig}
\usepackage{epstopdf}
\usepackage{dcolumn}
\usepackage{bm}
\usepackage{amsmath}
\usepackage{xspace}
\usepackage{multirow}
\usepackage{natbib}
\usepackage{color}

\begin{document}

\preprint{}

\title{Electron-Hole Asymmetry in Superconductivity of Pnictides Originated from the Observed Rigid Chemical Potential Shift }

\author{M. Neupane}
\affiliation{Department of Physics, Boston College, Chestnut Hill, MA 02467, USA}
\author{P. Richard}
\affiliation{WPI Research Center, Advanced Institute for Materials Research, Tohoku University, Sendai 980-8577, Japan}
\author{Y.-M. Xu}
\affiliation{Department of Physics, Boston College, Chestnut Hill, MA 02467, USA}
\author{K. Nakayama}
\affiliation{Department of Physics, Tohoku University, Sendai 980-8578, Japan}
\author{T. Sato}
\affiliation{Department of Physics, Tohoku University, Sendai 980-8578, Japan}
\author{T. Takahashi}
\affiliation{WPI Research Center, Advanced Institute for Materials Research, Tohoku University, Sendai 980-8577, Japan}
\affiliation{Department of Physics, Tohoku University, Sendai 980-8578, Japan}
\author{A. V. Federov}
\affiliation{Advanced Light Source, Lawrence Berkeley National Laboratory, Berkeley, CA 94720}
\author{G. Xu}
\affiliation{Beijing National Laboratory for Condensed Matter Physics, and Institute of Physics, Chinese Academy of Sciences, Beijing 100190, China}
\author{X. Dai}
\affiliation{Beijing National Laboratory for Condensed Matter Physics, and Institute of Physics, Chinese Academy of Sciences, Beijing 100190, China}
\author{Z. Fang}
\affiliation{Beijing National Laboratory for Condensed Matter Physics, and Institute of Physics, Chinese Academy of Sciences, Beijing 100190, China}
\author{Z. Wang}
\affiliation{Department of Physics, Boston College, Chestnut Hill, MA 02467, USA}
\author{G.-F. Chen}
\affiliation{Department of Physics, Renmin University of China,  Beijing 100872, China}
\author{N.-L. Wang}
\affiliation{Beijing National Laboratory for Condensed Matter Physics, and Institute of Physics, Chinese Academy of Sciences, Beijing 100190, China}
\author{H.-H. Wen}
\affiliation{Beijing National Laboratory for Condensed Matter Physics, and Institute of Physics, Chinese Academy of Sciences, Beijing 100190, China}
\author{H. Ding}
\affiliation{Beijing National Laboratory for Condensed Matter Physics, and Institute of Physics, Chinese Academy of Sciences, Beijing 100190, China}

\date{\today}
ÊÊÊÊÊÊÊÊÊÊÊÊ


\pacs{74.70.Pq, 71.30.+h, 79.60.-i}

\begin{abstract} 
We have performed a systematic photoemission study of the chemical potential shift as a function of carrier doping in a pnictide system based on BaFe$_2$As$_2$. The experimentally determined chemical potential shift is consistent with the prediction of a rigid band shift picture by the renormalized first-principle band calculations. This leads to an electron-hole asymmetry (EHA) in the Fermi surface (FS) nesting condition due to different effective masses for different FS sheets,  which can be calculated from the Lindhard function of susceptibility. This built-in EHA from the band structure, which matches well with observed asymmetric superconducting domes in the phase diagram, strongly supports  FS near-nesting driven superconductivity in the iron pnictides.
\end{abstract}

\maketitle
\pagebreak

Even though the detail of the pairing mechanism in the recently discovered iron-based superconductors is still under intense debate, several theoretical investigations \cite{Mazin, Kuroki, Wang, Seo, Cvetkovic} and experimental observations {\cite {Ding, Chris, RichardP, Terashima, Sekiba} strongly suggest the importance of inelastic  inter-band scattering between hole and electron Fermi surface (FS) pockets connected $via$ the antiferromagnetic (AF) wave vector. Within this framework, the pairing strength depends on near- or quasi-nesting, here defined as a large enhancement of the spin susceptibility at a well defined wave vector \cite{Cvetkovic}. The near-nesting conditions depend on the shape and size of the various FS pockets, which are tuned by the position of the chemical potential. The evolution of the chemical potential with carrier concentration is thus a key issue to understand the evolution of FS near-nesting and superconductivity in these materials.

The 122-structural phase of BaFe$_2$As$_2$ is particularly suitable for a systematic study of the chemical potential shift since it can be doped either by electrons or holes following Fe$^{2+}$$\rightarrow$ Co$^{3+}$ or Ba$^{2+}$$\rightarrow$ K$^{+}$ partial substitutions, respectively. Interestingly, the electron- and hole-doped sides of the phase diagram show some noticeable differences. For example, while the maximum $T_c$ value for the hole-doped side reaches 37 K at ambient pressure, it tops around 25 K for the electron-doped systems. Similarly, the superconducting dome extends to much higher doping in the hole-doped case, with an optimal concentration of around 0.2 hole/Fe against 0.08 electron/Fe for the electron-doped side. 

Angle-resolved photoemission spectroscopy (ARPES) is a powerful tool to access directly the electronic structure with respect to the chemical potential. Our previous ARPES studies have already revealed strong variations in the pairing strengths associated with the various FS sheets in the electron-doped compounds \cite {Terashima} as compared to the hole-doped ones \cite {Ding, Nakayama}, as well as the deterioration of the near-nesting conditions in highly overdoped samples for which $T_c$ vanishes or is significantly suppressed \cite{Sekiba, Sato}. Although near-nesting was naturally proposed to explain these anomalies, this concept has not been linked to the origin of the electron-hole asymmetry (EHA) and up to date there is still no systematic investigation of the impact of  the chemical potential shift on the band structure throughout the whole phase diagram.  

In this letter, we present a systematic ARPES study of the chemical potential as a function of carrier doping in the 122-pnictides. With doping, the chemical potential moves smoothly with respect to the low-energy valence states, in agreement with our local density approximation (LDA) calculations. 
However, we observed anomalously larger (smaller) core level shift than the valence band shift on the hole (electron)-doped side for the relatively undisturbed As 3$d$ core levels, possibly due to the screening effect which increases (decreases) the core level shifts upon hole (electron) doping.
Based on a rigid band shift approximation justified by our experimental results, we computed the doping dependence of the Lindhard spin susceptibility at the AF wave vector, and found that the Lindhard function itself is asymmetric as a function of doping, in a similar fashion as the asymmetry between the hole and electron superconducting domes. This strongly supports FS-near-nesting-enhanced superconductivity in the pnictides.
 
\begin{figure}[htbp]
\begin{center}
\includegraphics[width=3.5in]{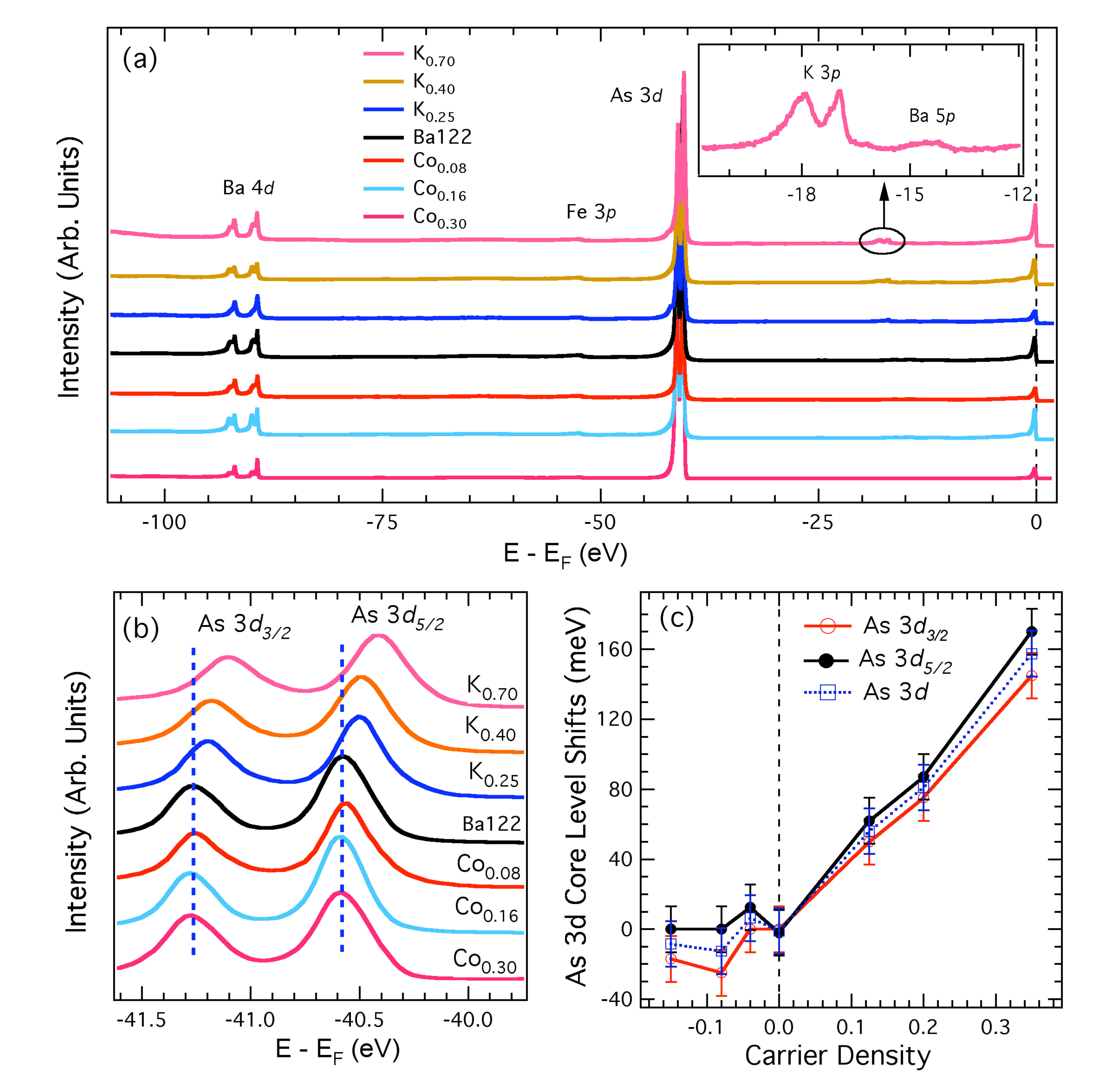}
\caption{\label{Figure1} (a) Core levels of the Ba122 series recorded with a photon energy of 140 eV. The inset shows a zoom of the core levels of the K$_{0.70}$ sample in the 12-21 eV  binding energy range. (b) Zoom of the As 3$d$ core levels. (c) Doping dependence of the As 3$d_{3/2}$ and  As 3$d_{5/2}$ core level energies as a function of carrier density (half of the x value). The average is represented by the blue dotted line.}
\end{center}
\end{figure}

The high-quality single crystals of the 122 series used in this study were grown by the flux method \cite {Chen}. Low-energy electron diffraction on mirror-like cleaved surfaces show a sharp 1 x 1 pattern in the non-magnetic phase. High-resolution (4-20 meV) ARPES measurements of the low-energy electronic structure were performed in the photoemission laboratory of Tohoku University using a microwave-driven Helium source ($hv$ = 21.218 eV) and core level studies were done at the Synchrotron Radiation Center and the  Advanced Light Source in USA, as well as at the Photon Factory in Japan, using various photon energies. Our experiments have been performed  using  high-efficiency VG-Scienta SES-100, SES-2002 and R4000 electron analyzers. Samples were
cleaved {\it in situ} and measured at 7-40 K in a vacuum better than 1 x 10$^{-10}$ torr. The samples have been found to be very stable and without degradation for
the typical measurement period of 20 hours.

Photoemission allows measurement of the core level states relative to the chemical potential. It has been used widely in the past to study the chemical potential shift in high-$T_c$ cuprates \cite{Yagi, Harima, Golden}. 
Fig.~1(a) shows a comparison of the core levels in the 0-110 eV binding energy range of 7 samples distributed in the electron-doped and hole-doped sides of the phase diagram. These samples are  BaFe$_{1.70}$Co$_{0.30}$As$_2$ ($T_c$ = 0 K),  BaFe$_{1.84}$Co$_{0.16}$As$_2$ ($T_c$ = 20 K), BaFe$_{1.92}$Co$_{0.08}$As$_2$ ($T_c$ = 0 K), Ba$_{2}$Fe$_2$As$_2$ ($T_c$ = 0 K),  Ba$_{0.75}$K$_{0.25}$Fe$_2$As$_2$ ($T_c$ = 26 K), Ba$_{0.60}$K$_{0.40}$Fe$_2$As$_2$ ($T_c$ = 37 K), and Ba$_{0.30}$K$_{0.70}$Fe$_2$As$_2$ ($T_c$ = 22 K). For simplicity, here after we call them  Co$_{0.30}$, Co$_{0.16}$, Co$_{0.08}$, Ba122, K$_{0.25}$, K$_{0.40}$ and K$_{0.70}$, respectively. From low to high binding energies, we observed the Fe 3$d$ (around the Fermi level), Ba 5$p$ ($\sim$14.5 eV), K 3$p$ ($\sim$18 eV), As 3$d$ ($\sim$ 40.4 and 41.3 eV) and Ba 4$d$ ($\sim$ 89.5 and 92 eV) states, respectively. In particular, the As 3$d$ peaks are very strong in all compounds regardless of Co and K doping. Based on a previous photoemission study \cite{Golden1} that indicates that the As atoms in  BaFe$_{2}$As$_2$ are not perturbed significantly at the cleaved surface, we used the As 3$d$ core levels to investigate the doping dependence of the chemical potential. In Fig.~1(b), we show a zoom of the As 3$d$ core levels of all compounds. The position of the peaks moves towards the lower binding energies  as K concentration increases. In contrast, the peak positions are almost unaffected by Co-doping. We plot in Fig.~1(c) the shift of the  As 3$d_{3/2}$ and As 3$d_{5/2}$ levels as a function of carrier density, which is half of the value x for both K and Co dopings. The blue dashed line in Fig.~1(c) gives the average of the As 3$d$ peaks as a function of doping.

\begin{figure}[htbp]
\begin{center}
\includegraphics[width=3.5in]{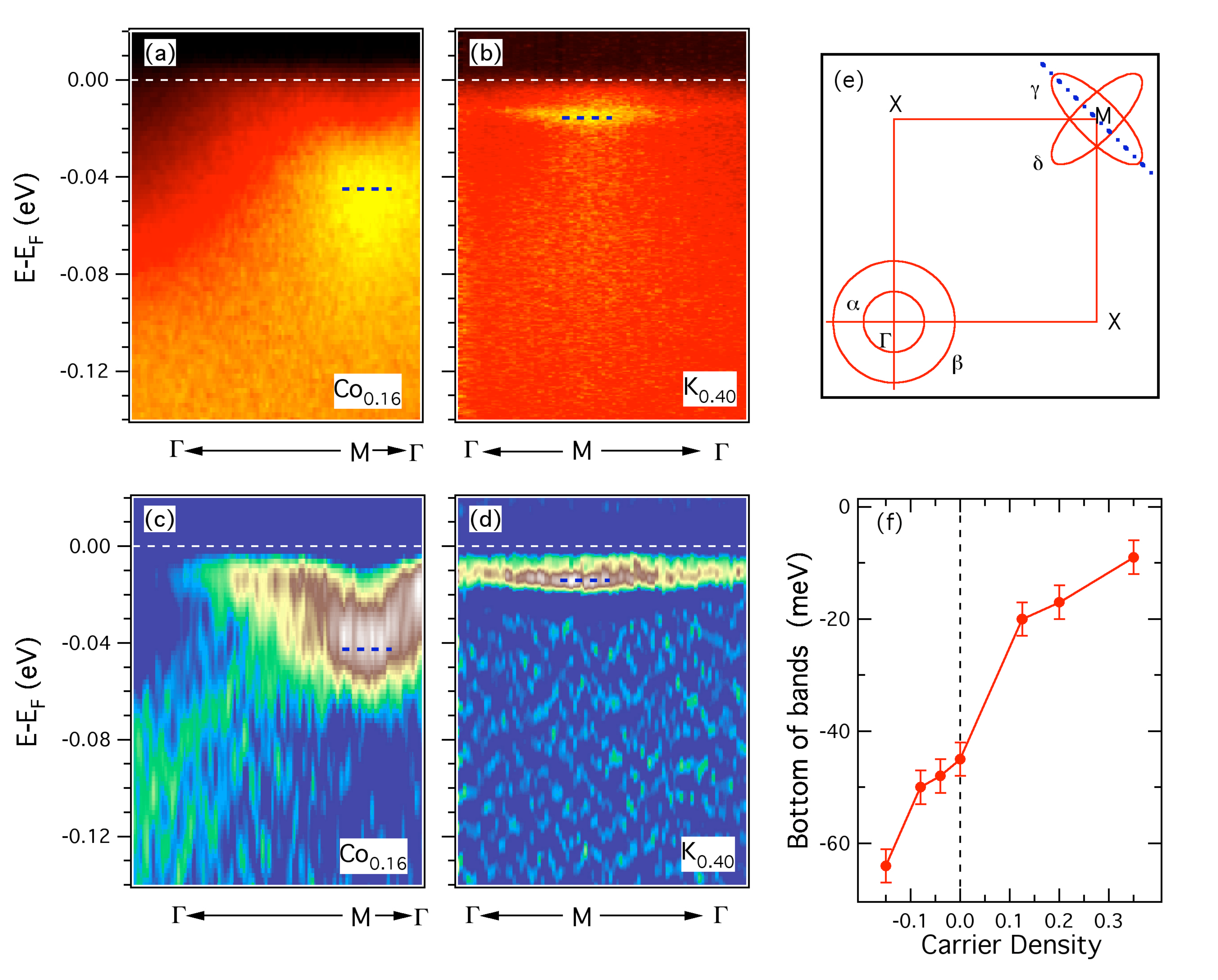}
\caption{\label{Figure 2} (a) and (b) ARPES intensity plots of the Co$_{0.16}$ and K$_{0.40}$ samples, respectively, along the cut passing through M indicated by a dashed line in panel (e).
(c) and (d) Corresponding second derivative intensity plots.
 (f) Bottom of the electron bands 
 versus carrier density.}
\end{center}
\end{figure}

An alternative and more direct determination of the chemical potential shift is obtained by looking at the band dispersion near the Fermi level ($E_F$).  In Figs.~2(a) and (b), we present ARPES intensity plots of the Co$_{0.16}$ and K$_{0.40}$ samples along a cut passing through  M as indicated in Fig.~2(e). The corresponding second derivative intensity plots are displayed in Figs.~2(c)-(d). The blue dashed lines are guides to the eye indicating the bottom of the upper electron band (the $\gamma$ band as defined in Ref. \onlinecite{Hong1}).
The bottom of this electron band at the M point  moves down from $E_F$ as the signed concentration decreases (more electrons), which is what we expect from simple band filling. In particular, this behavior supports the assumption that the Fe $\rightarrow$ Co substitution electron-dope the Fe layer, in contrast to a recent density functional theory calculation suggesting that Co and Ni only act as scattering centers in the Fe planes \cite{wadati}. It is also consistent with the observation of a downshift of the $\Gamma$-centered holelike bands in the Co-doped side \cite{Sekiba}. 
 Fig.~2(f) summarizes our results of the seven differently doped samples and gives the position of the bottom of the electron band as a function of the carrier density.

\begin{figure}[htbp]
\begin{center}
\includegraphics[width=3.5in]{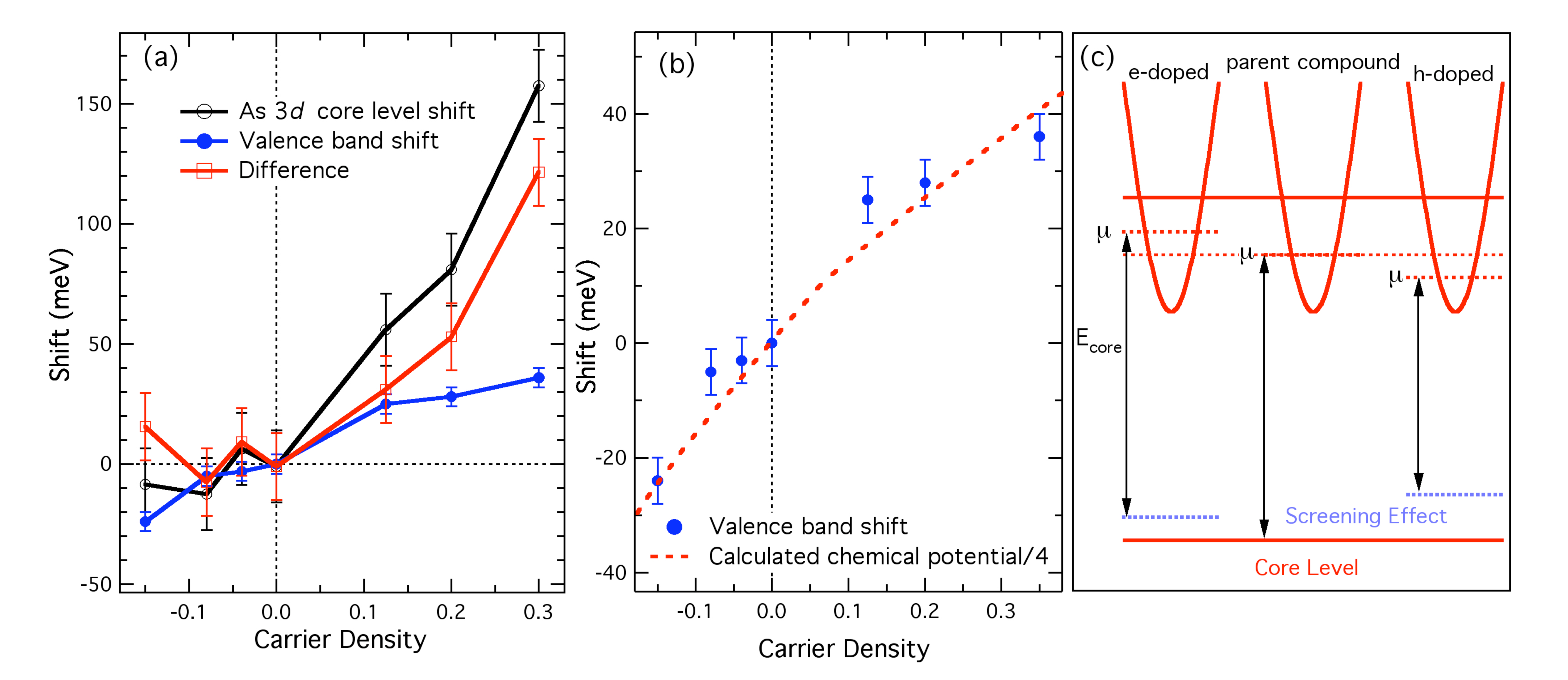}
\caption{\label{Figure 3} {Valence band and core level shifts as a function of the carrier density.} (a) The black line with open circles shows the core level shifts while the blue line with closed circles gives the valence band shifts measured from the band bottom. The red line with open squares gives the difference between the core level and chemical potential shifts. (b) The blue dots are the valence band shifts shown in panel a, and the red dash line is the LDA calculated values of the chemical potential divided by a factor of 4. (c) Pictorial representation of  the explanation of the core level and the chemical potential shifts as a function of carrier density.}
\end{center}
\end{figure}

At this point, it is instructive to compare LDA calculations to the core level shifts and the shift of the bottom of the electron band, which corresponds to the chemical potential shift in a rigid band picture. The results are summarized in Fig.~3. It is clear from Fig.~3(a) that the  core level shift is not the same as the shift of the valence band, and the difference between them are larger on the hole-doped side, which will be discussed below.  Interestingly, the theoretically calculated chemical potential shift is very much consistent with the observed valence band shift when theoretical values are divided by 4 as shown in Fig.~3(b), which is understood in terms of the band renormalization reported in previous ARPES studies \cite {Hong1, Terashima, Sekiba, Sato}. This indicates that the shift of the valence band corresponds to the chemical potential shift, and consequently, the rigid band picture derived from the renormalized band structure is valid. 

The  core level shift can be understood as follows. 
The  core level shift $\Delta$$E$ is related to the chemical potential shift $\Delta\mu$ by the relation:
\begin{equation}
\notag
\Delta E = -\Delta\mu+ K\Delta Q + \Delta V_M + \Delta E_R
\end{equation}
where $\Delta$$Q$  is the change in valency, $K$ is a constant, $\Delta$$V_M$  is a shift due to change in the Madelung potential, and  $\Delta$$E_R$ is the change in the core-carrier screening \cite{Hufner}. Doping is not expected to change the As valency. This implies that the term $K$$\Delta$$Q$ can be neglected. Therefore, the difference between the core level and the chemical potential shift represented in Fig.~3(a) by the red line is only related to  $\Delta$$V_M$  and  $\Delta$$E_R$. It is known that the screening term $\Delta$$E_R$ is proportional to the mobile carrier concentration, thus one expects that it has the same sign on the electron- and hole-doped sides and increases with doping. Such doping dependence of the screening term, as indicated in Fig.~3(c), will increase (reduce) the core level shift caused by the chemical potential shift on the hole (electron)-doped side. This is consistent with our observation of different behaviors of the core level shift on hole- and electron-doped sides. We note that the change of the Madelung term $\Delta$$V_M$ may not be same on hole- and electron-doped sides, which can further enhance the difference of the core level shift on the two sides.

\begin{figure}[htbp]
\begin{center}
\includegraphics[width=3in]{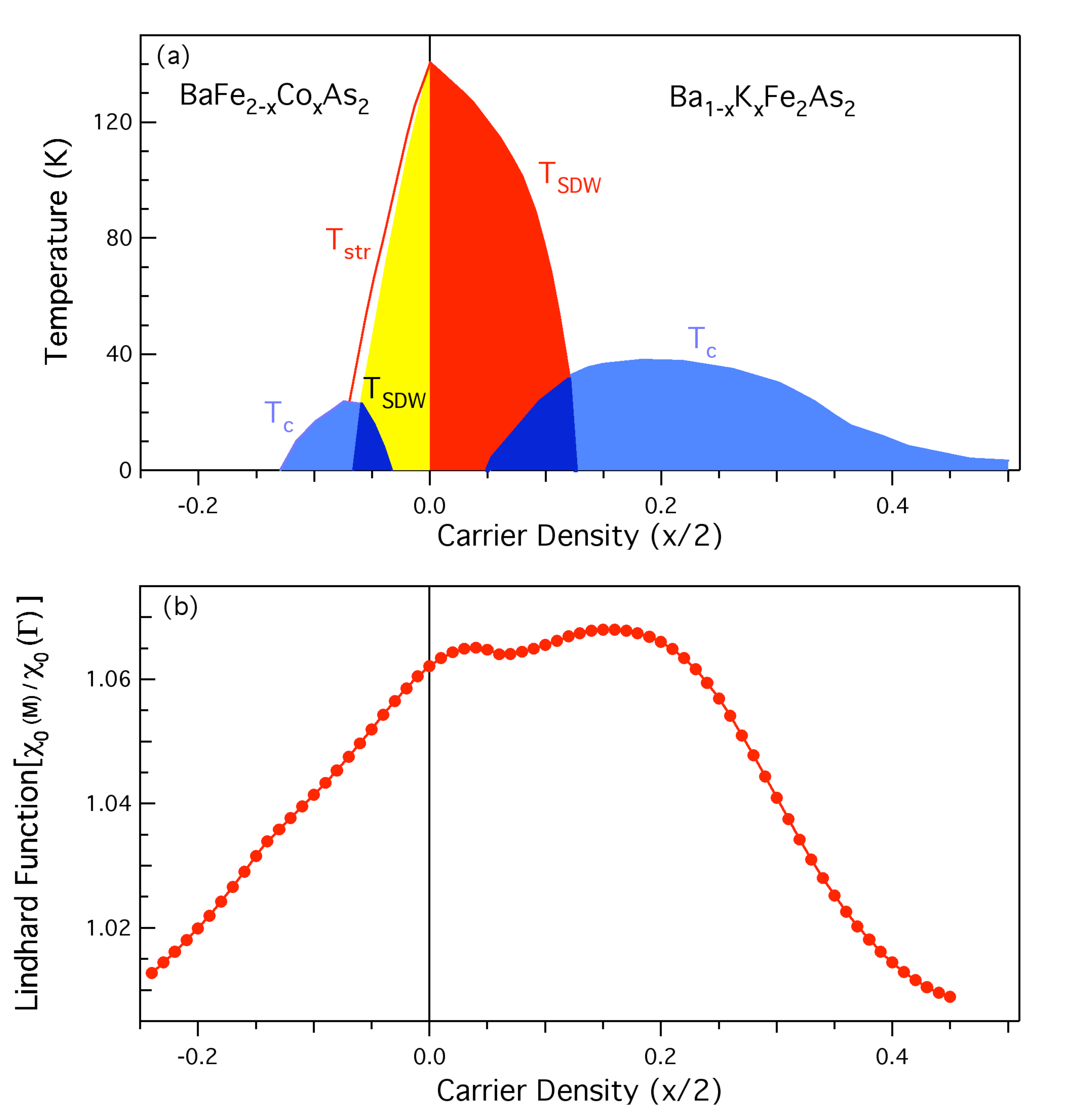}
\caption{\label{Figure 4} { (a) Phase diagram of the hole- and electron-doped Ba-122 systems taken from references \cite{rotter} and \cite{Canfield}, respectively. $T_c$, $T_{SDW}$ and $T_{str}$ refer to the superconducting, the SDW and the tetragonal to orthorombic structural transitions, respectively. (b) Doping dependence of the Lindhard function at the M point (near-nesting wave vector) normalized by its value at the zone center. The Lindhard function was obtained by using LDA calculation.}}
\end{center}
\end{figure}

The above analysis suggests that a rigid band picture constitutes a good first approximation of the evolution of the chemical potential in the 122 family of iron-pnictides. We now ask a simple but fundamental question: is FS near-nesting able to explain the electron-hole asymmetry of the superconducting domes shown in the phase diagram of the 122-pnictides of Fig.~4(a)?
To answer this question, it is necessary to compute the spin susceptibility. It is especially important to understand how the susceptibility evolves at the near-nesting (or AF) wave vector. We use the band structure calculated by LDA to compute the doping dependence of the Lindhard spin susceptibility at the near-nesting wave vector \cite{Xu_LDA, Wang_LDA}. We limit our calculations to the elastic component of the spin susceptibility. The results are displayed in Fig.~4(b). Interestingly, the hole- and electron-doped sides exhibit a strong asymmetry: while the Lindhard function decreases monotonically on the electron-doped side (with a small shoulder around $\sim$0.12), it keeps a high value for a wide hole doping range before starting to decrease. Remarkably, the maximum value of the calculated susceptibility is obtained near the experimental optimal hole doping, and the Lindhard function using the FS nesting wave vector tracks the superconducting transition qualitatively well. 
It is important to note that such an asymmetry in the Lindhard susceptibility would lead to a higher dielectric function and consequently to a larger screening effect on the hole-doped side. This is qualitatively consistent with the observed larger difference between the core level shift and the chemical potential shift on the hole-doped side shown in Fig.~3(a).
We caution that the non-magnetic LDA calculations are no longer valid in the spin density wave (SDW) state because the band structure undergoes unconventional band folding that leads to the formation of Dirac cones \cite{Richard}. 

The basic reason for electron-hole asymmetry in the calculated Lindhard function is as follows. The effective masses of the holelike bands, especially the $\beta$ band, are larger than that of the electronlike bands at the M point, as observed by ARPES \cite {Hong1} and quantum oscillation experiments \cite{dHvA}. To satisfy the Luttinger theorem, their top of band at zero doping must thus be closer to $E_F$ than the bottom of the electron bands. Indeed, even for optimally hole-doped samples, the top of the $\alpha$ band is located only 25 meV above $E_F$  \cite{RichardP}. As a consequence, the holelike bands sink below $E_F$ with electron doping much faster than the bottom of the electron bands are pushed above $E_F$ with hole doping. Therefore, the FS near-nesting conditions are more robust in the hole-doped case.
The built-in asymmetry regarding the FS near-nesting condition on the electron- and hole-doped sides offer a simple but powerful clue that the FS near-nesting with the AF wave vector triggers superconductivity in the pnictides.

In conclusion, we have presented the doping dependence of  the chemical potential in the 122 family of iron-pnictides. As a first approximation, our results are consistent with a rigid band shift and with renormalized LDA calculations. The doping dependence of the As 3$d$ core levels does not follow that of the chemical potential, suggesting a non-negligible screening effect. Within the rigid band shift approximation, the calculated Lindhard function at the FS-nesting wave vector based on the LDA band structure reveals an electron-hole asymmetry in the iron pnictides, which matches well with the observed electron-hole asymmetry of the superconducting domes in the phase diagram. Our findings reveal the importance of FS near-nesting in the pairing mechanism of the iron-based superconductors.

This work was supported by grants of US DMR-0800641,
DMR-0704545, US DOE DE-SC0002554, and China NSF. The Tohoku group acknowledge support from JSPS, JST-TRIP,  JST-CREST and MEXT of Japan. This work is
based upon research conducted at the Synchrotron Radiation Center
supported by NSF DMR-0537588,  the Advanced Light Source supported by DOE No. De-AC02-05CH11231 and at KEK-PF under the approval of Photon Factory Program Advisory Committee Proposal No. 2009S2-005 at the Institute of Material Structure Science, KEK.


\end{document}